\documentclass[twocolumn]{aastex63}
\usepackage{times}
\usepackage{amsmath}
\usepackage{textcomp}
\usepackage{amssymb}
\usepackage{natbib}
\usepackage{float}
\usepackage{color}
\usepackage{graphicx}
\usepackage{epstopdf}
\newcommand{\lum}{erg\,s$^{-1}$}
\newcommand{\fermi}{{\it Fermi}}

\newcommand{\swift}{{\it Swift}}

\newcommand{\gm}{$\gamma$}
\received{\today}
\revised{\today}
\accepted{\today}
\submitjournal{ApJS}

\shorttitle{3C 279 and 3C 454.3}
\shortauthors{Paliya et al.}

\begin{document}

\title{On the Origin of Gamma-ray Flares from Bright \textbf{\textit {Fermi}} Blazars}

\correspondingauthor{Vaidehi S. Paliya}
\email{vaidehi.s.paliya@gmail.com}

\author[0000-0001-7774-5308]{Vaidehi S. Paliya}
\affiliation{Aryabhatta Research Institute of Observational Sciences (ARIES), Manora Peak, Nainital 263001, India}

\author[0000-0002-8434-5692]{M. B\"ottcher}
\affiliation{Centre for Space Research, North-West University, Potchefstroom, 2531, South Africa}

\author[0000-0003-0685-3621]{Mark Gurwell}
\affiliation{Center for Astrophysics, Harvard \& Smithsonian, Cambridge, MA 02138, USA}

\author[0000-0002-4998-1861]{C. S. Stalin}
\affiliation{Indian Institute of Astrophysics, Block II, Koramangala,Bengaluru, Karnataka 560034, India}

\begin{abstract}
The origin of \gm-ray flares observed from blazars is one of the major mysteries in jet physics. We have attempted to address 
this problem following a novel spectral energy distribution (SED) fitting technique that explored the flaring patterns identified in the broadband 
SEDs of two \gm-ray bright blazars 3C 279 ($z=0.54$) and 3C 454.3 ($z=0.86$), 
using near-simultaneous radio-to-\gm-ray observations. For both sources, the 
\gm-ray flux strongly correlates with the separation of the SED peaks and the Compton dominance. We propose that spectral hardening 
of the radiating electron population and/or enhancement of the Doppler factor can naturally explain these observations. In both cases, magnetic reconnection 
may play a pivotal role in powering the luminous \gm-ray flares.

\end{abstract}


\section{Introduction}{\label{sec:Intro}}
The \gm-ray emission from flat-spectrum radio quasars (FSRQs) offers a powerful diagnostic tool to probe the innermost regions of jets and their 
surroundings. FSRQs are a special class of active galactic nuclei (AGN) having relativistic jets oriented close to the line of sight to the observer and 
characterized by broad optical emission lines. The typical spectral energy distribution (SED) of an FSRQ is dominated by the non-thermal jet emission 
and is characterized by two broad bumps. In the widely accepted leptonic emission scenario, the low-energy bump produced by the synchrotron process 
peaks in the sub-millimeter (mm) or the infrared (IR)-optical energy regime. The high energy bump, on the other hand, produced by inverse Compton 
scattering, peaks in the MeV range, leading to a steep \gm-ray spectrum in the GeV energy band \citep[e.g.,][]{2020ApJ...892..105A}. These intriguing 
objects exhibit flux variations at all accessible wavelengths with the most extreme variability, on the timescale of hours or less, observed at GeV energies
\citep[see, e.g.,][]{2011A&A...530A..77F}. During high-activity states, many FSRQs show a rising $\gamma$-ray spectrum in $\nu F_{\nu}$, indicating a 
shift of the inverse Compton peak from MeV to GeV energies \citep[cf.][]{2016ApJ...817...61P}. In a few cases, the associated synchrotron peak is also 
observed to shift from far-IR to optical-UV energies \citep[e.g.,][]{2019A&A...627A.140A}. On the other hand, there are observations indicating that only 
the inverse Compton peak shifted to higher frequencies without any significant change in the synchrotron peak \citep[][]{2016ApJ...817...61P}. Such 
contrasting behaviors indicate the complex interplay between particle acceleration and radiation mechanisms in relativistic jets and, at the same time, 
provide tantalizing clues to explore the underlying causes responsible for the flux variability. The key to understanding the origin of \gm-ray flares lies 
in unraveling the spectral patterns shown by FSRQs across the electromagnetic spectrum. 

\begin{figure*}[t!]
\hbox{\centering
\includegraphics[scale=0.24]{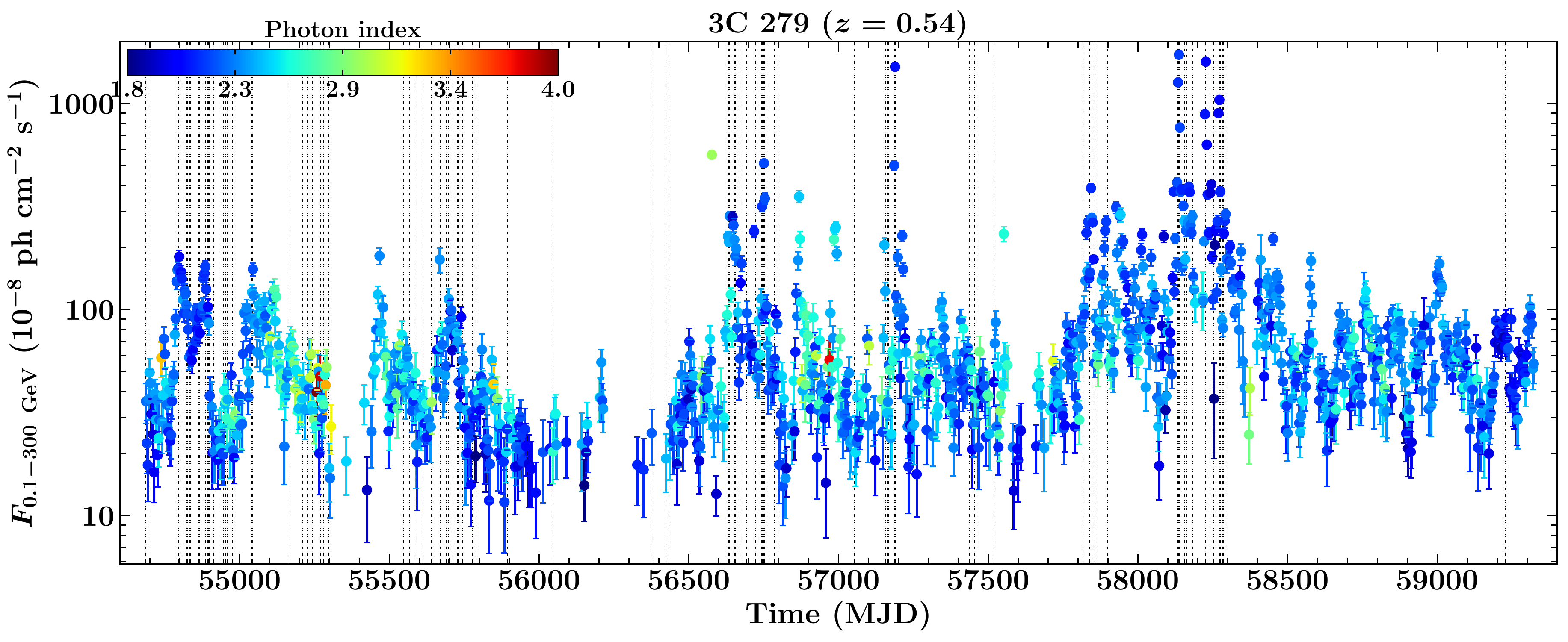} 
\hspace{0.1cm}
\includegraphics[scale=0.24]{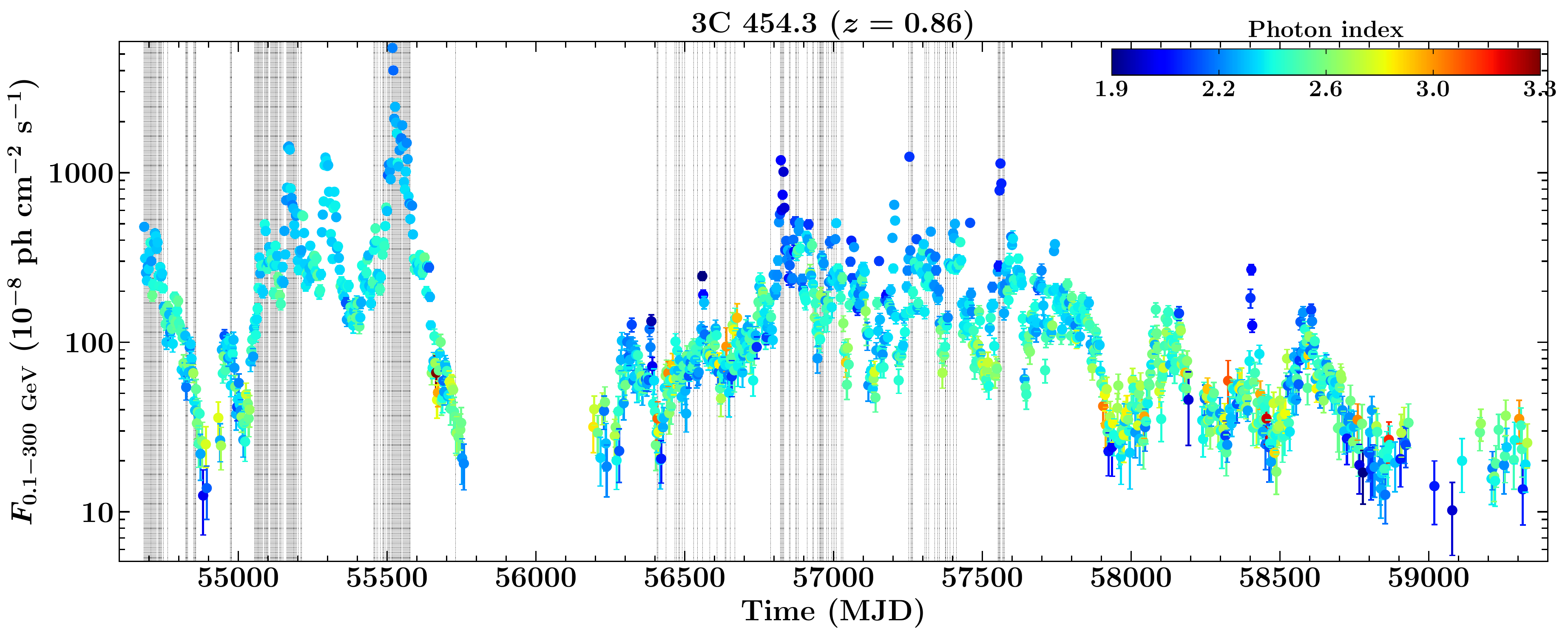} 
}
\caption{The 3-day binned \gm-ray light curves of 3C 279 (left) and 3C 454.3 (right). Vertical dotted lines refer to the epochs 
when there is at least one simultaneous observation available each in the radio, NIR-to-UV, X- and \gm-ray bands.}
\label{fig:lc}
\end{figure*}

We have attempted to address this outstanding problem by carrying out a systematic study of the broadband SEDs of two well-known \gm-ray detected 
blazars, namely 3C 279 ($z=0.54$) and 3C 454.3 ($z=0.86$). The motivation is to understand the evolution of the low- and high-energy SED peaks as 
a function of the \gm-ray activity following a novel SED fitting approach which treated the SED peak transitions separately. We further developed a theoretical 
model to decipher the observed patterns revealing the origin of \gm-ray flares. We selected 3C 279 and 3C 454.3 because these are among the brightest 
\gm-ray blazars with hundreds of existing multi-wavelength observations supplemented with more-than-a-decade of continuous monitoring with the 
\fermi-Large Area Telescope (\fermi-LAT). The dense sampling has allowed us to study their multi-epoch SED evolution using contemporaneous data 
covering all accessible electromagnetic bands. Moreover, near-IR (NIR) to optical-ultraviolet (UV) spectra of both objects are 
strongly dominated by non-thermal synchrotron emission from the jet, with signatures of direct accretion disk emission only emerging in very low
activity states \citep[e.g.,][]{Pian1999} and at frequencies far beyond the synchrotron peak. Thus, we are able to 
reliably estimate the location of the synchrotron peak based on the observed NIR -- UV spectra. The following cosmology parameters were adopted: 
$H_0=67.8$~km~s$^{-1}$~Mpc$^{-1}$, $\Omega_m = 0.308$, and $\Omega_\Lambda = 0.692$ \citep[][]{2016A&A...594A..13P}

\section{Data Reduction and Analysis}\label{sec:data_red}

\begin{figure*}[t!]
\hbox{\centering
\includegraphics[scale=0.356]{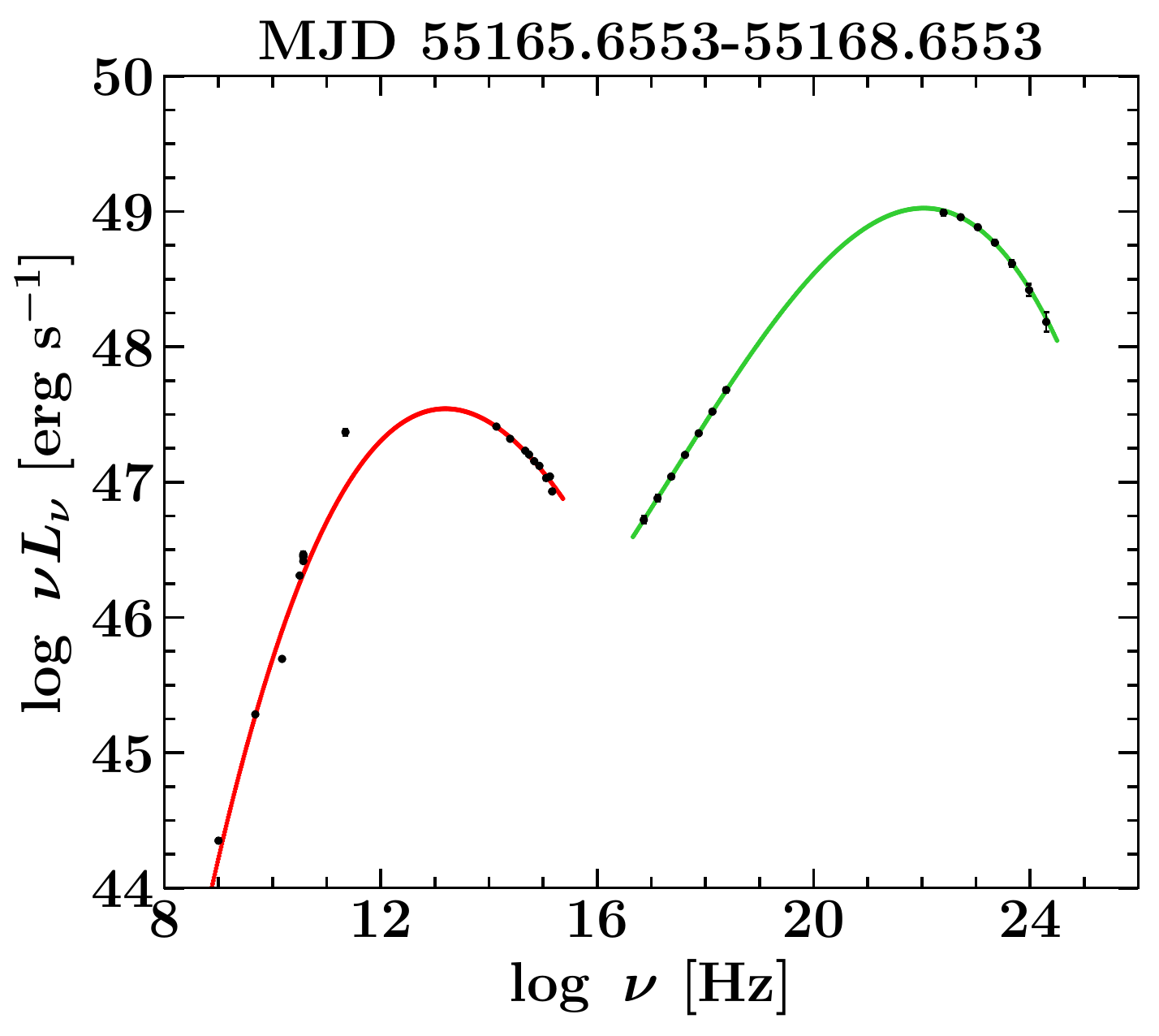} 
\includegraphics[scale=0.32]{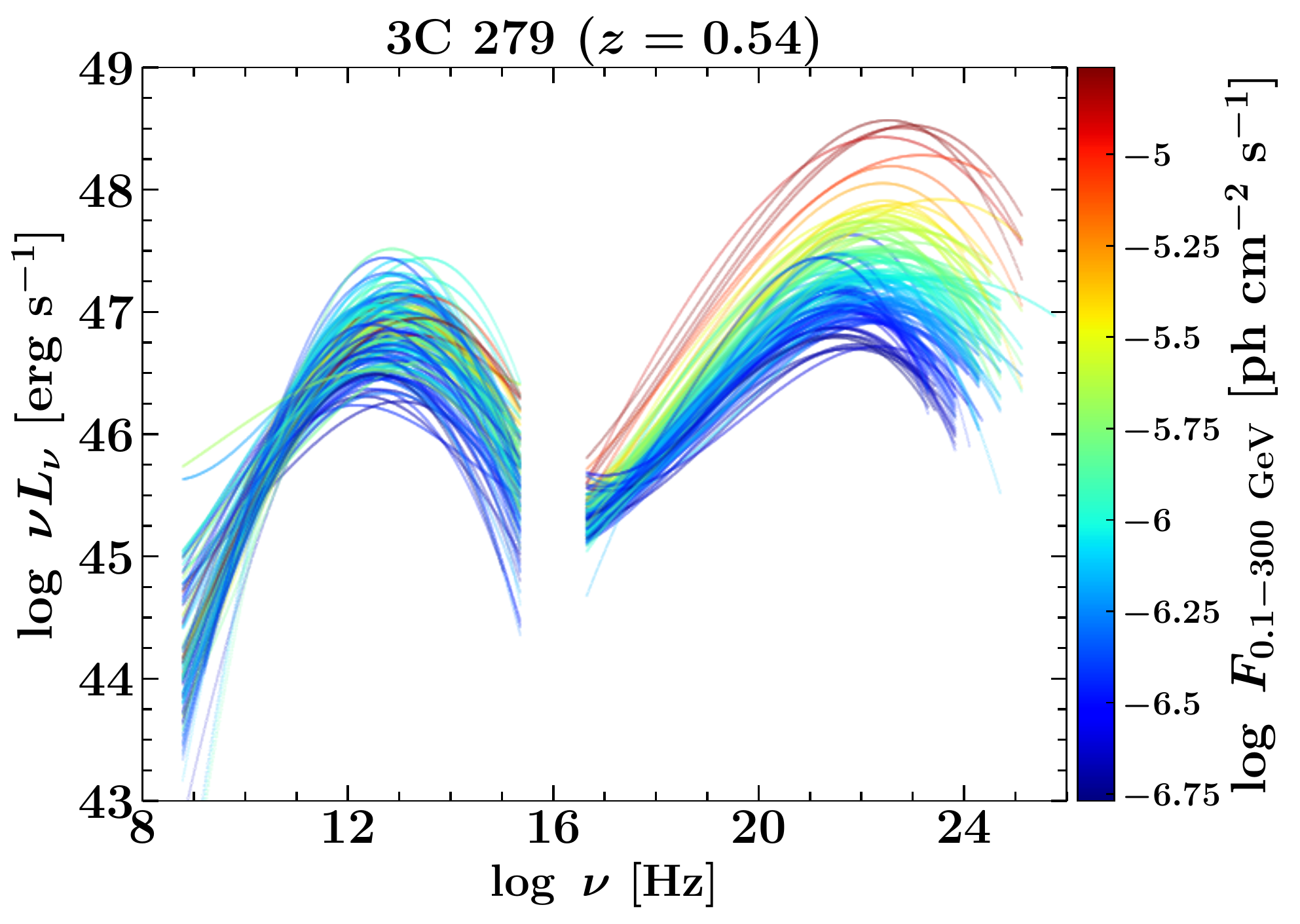} 
\includegraphics[scale=0.32]{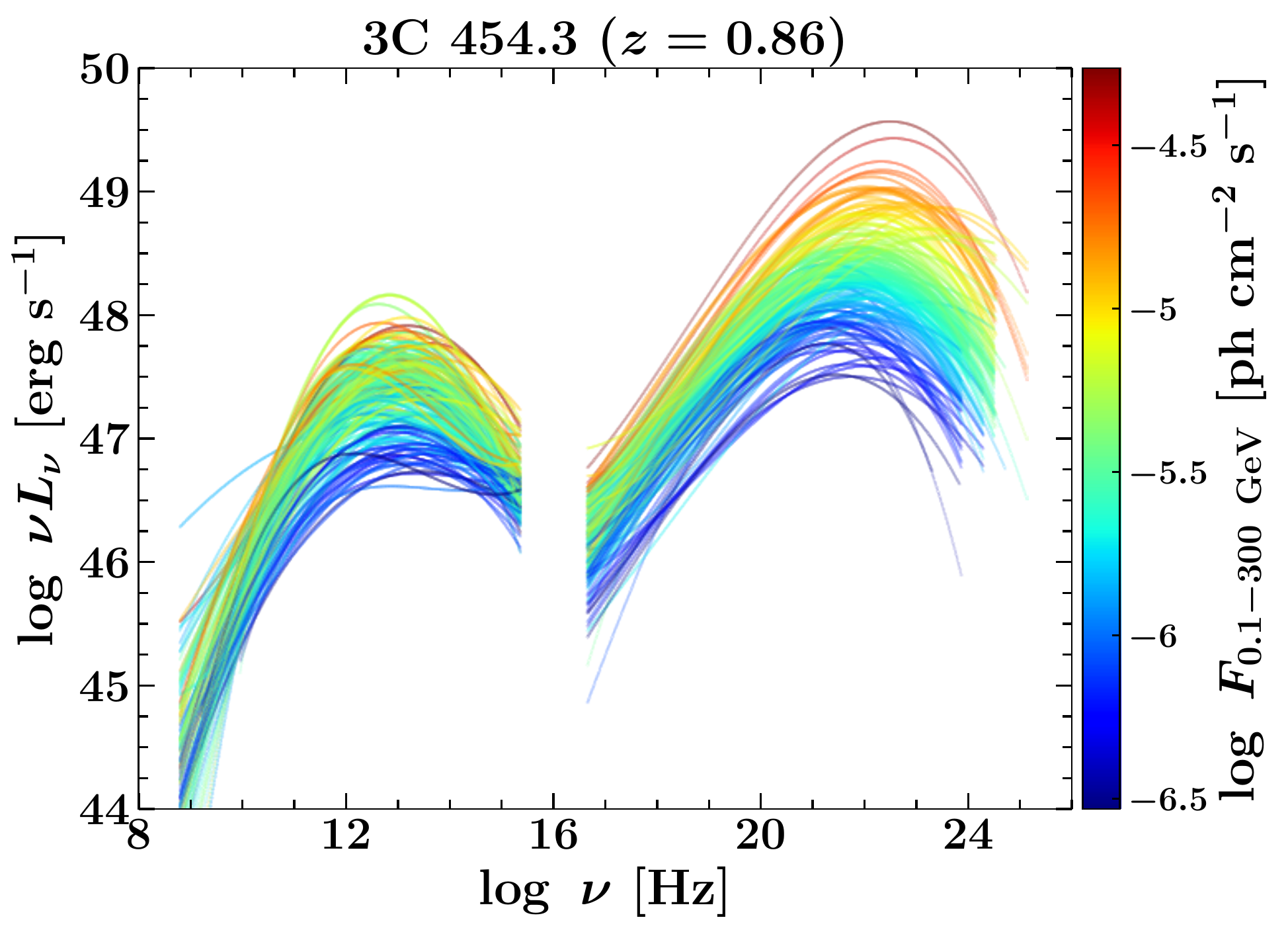} 
}
\caption{An example of fitting the SED with third-order polynomials is demonstrated in the left panel. The middle and 
right panels show all of the fitting results for 3C 279 and 3C 454.3, respectively, with the color coding done based on the \gm-ray photon flux level.}
\label{fig:sed_fit}
 \end{figure*}
 
{\it Gamma-rays}: We analyzed 0.1$-$300 GeV data taken with the \fermi-LAT between 2008 August 4 and 2021 May 5 (MJD 54683-59339) to generate 
3-day binned \gm-ray light curves and spectra. We followed the standard data reduction 
procedure\footnote{http://fermi.gsfc.nasa.gov/ssc/data/analysis/documentation/} fully described in various works \citep[cf.][]{2019ApJ...871..211P}
and we briefly elaborate it here. A 15$^{\circ}$ region of interest (ROI) was chosen to select SOURCE class events which were subjected to the relational filter {\tt `DATA\_QUAL$>$0 and LAT\_CONFIG==1'} to determine good time intervals. Additionally, we adopted a zenith angle cut of $z_{\rm max}<90^{\circ}$ to avoid contamination from Earth's albedo. A binned likelihood fitting technique was used to determine the spectral parameters of sources of interest. For this purpose, we considered all \gm-ray sources located within 20$^{\circ}$ from the center of the ROI and included in the fourth catalog of the \fermi-LAT detected sources \citep[4FGL-DR2;][]{2020ApJS..247...33A,2020arXiv200511208B}. The standard Galactic and isotropic background templates\footnote{https://fermi.gsfc.nasa.gov/ssc/data/access/lat/BackgroundModels.html} were also used. The spectral parameters of all sources lying within the ROI were kept free in the likelihood fit, whereas, those lying outside ROI were fixed to their 4FGL-DR2 values. The significance of the \gm-ray detection was quantified using maximum likelihood test statistic TS=2$\Delta \log (\mathcal{L})$ where $\mathcal{L}$ is the ratio of the likelihood values for models with and without a \gm-ray point object. We first performed a likelihood fit for the whole duration covered in this work and then froze the spectra parameters of all sources whose TS values were found to be $<$25. This updated sky model was then used to generate 3-day binned light curve and spectra. In this study, we considered only those epochs during which the target quasars were significantly detected ($\gtrsim$5$\sigma$) in the \gm-ray band.  The generated light curves are shown in Figure~\ref{fig:lc}.
 
{\it X-rays}: We used the ``\swift-XRT data products generator\footnote{http://www.swift.ac.uk/user\_objects/}" \citep[][]{2009MNRAS.397.1177E} 
to extract 0.3$-$10 keV spectra from the observations taken with the {\it Neil Gehrels} \swift~X-ray telescope in the time interval MJD 54683 -- 59339. Based on the current count rate, this tool automatically takes into account the possible pile-up effects and choose the source and background regions of appropriate sizes. An 
absorbed power-law model was adopted to derive the source flux for each observation using the Galactic neutral hydrogen column density from \citet[][]{2005AA...440..775K}. The spectral fitting was carried out in XSPEC \citep[][]{1996ASPC..101...17A}.

{\it NIR-UV}: Data from the \swift~Ultraviolet and Optical telescope (UVOT --- $V, B, U, W1, M2$ and $W2$ filters), Steward observatory 
\citep[$V$ and $R$ bands;][]{2009arXiv0912.3621S} and Small and Medium Aperture Research Telescope System \citep[SMARTS --- $B, V, R, J$ 
and $K$ filters;][]{2012ApJ...756...13B} were used to extract the source magnitudes in the NIR, optical, and UV bands. They were corrected for the 
Galactic extinction following \citet[][]{2011ApJ...737..103S} and converted to flux units using the standard zero points 
\citep[][]{1998A&A...333..231B,2011AIPC.1358..373B}.

{\it Radio}: We considered the multi-frequency radio fluxes made available by the University of Michigan Radio Astronomy Observatory 
\citep[UMRAO --- 4.8, 8, and 14.5 GHz;][]{1985ApJS...59..513A}, the Submillimeter Array\footnote{http://sma1.sma.hawaii.edu/callist/callist.html} 
\citep[230 and 350 GHz;][]{2007ASPC..375..234G}, and the Atacama Large Milimeter/Submilimeter Array 
\citep[ALMA --- 35 GHz to 950 GHz;][]{2019MNRAS.485.1188B}. The radio fluxes published in \citet[][]{2019A&A...626A..60A} and 
\citet[][]{2020MNRAS.492.3829L} were also used.
 
In the appendix, we provide the output of the multiwavelength data reduction in Table~\ref{tab1} and \ref{tab2} for 3C 279 and 3C 454.3, respectively. 
\section{Evolution of the SED Peaks}\label{sec:sed_peak}

Our primary objective is to study the evolution of the SED peaks as a function of the \gm-ray activity of 3C 279 and 3C 454.3. Therefore, from 
the \gm-ray light curves, only those epochs were chosen that have at least one observation each in the radio and NIR-UV (covering the synchrotron 
peak), and X-ray and \gm-ray (constraining the inverse Compton peak) bands. This criterion of selecting near-simultaneous data across the 
electromagnetic spectrum has led to the identification of 155 and 149 epochs for 3C 279 and 3C 454.3, respectively (Figure~\ref{fig:lc}). To 
determine the peak frequencies and luminosities, we fitted a third-order polynomial following the Markov Chain Monte Carlo technique 
implemented in the software {\tt emcee} \citep[][]{emcee}. Both SED peaks were fitted independently and we derived the uncertainties 
based on the 16th, 50th, and 84th percentiles of the sample in the marginalized distribution. We show the results of this fitting procedure 
in Figure~\ref{fig:sed_fit}.

\begin{figure*}[t!]
\vbox{\centering
\hbox{
\includegraphics[scale=0.36]{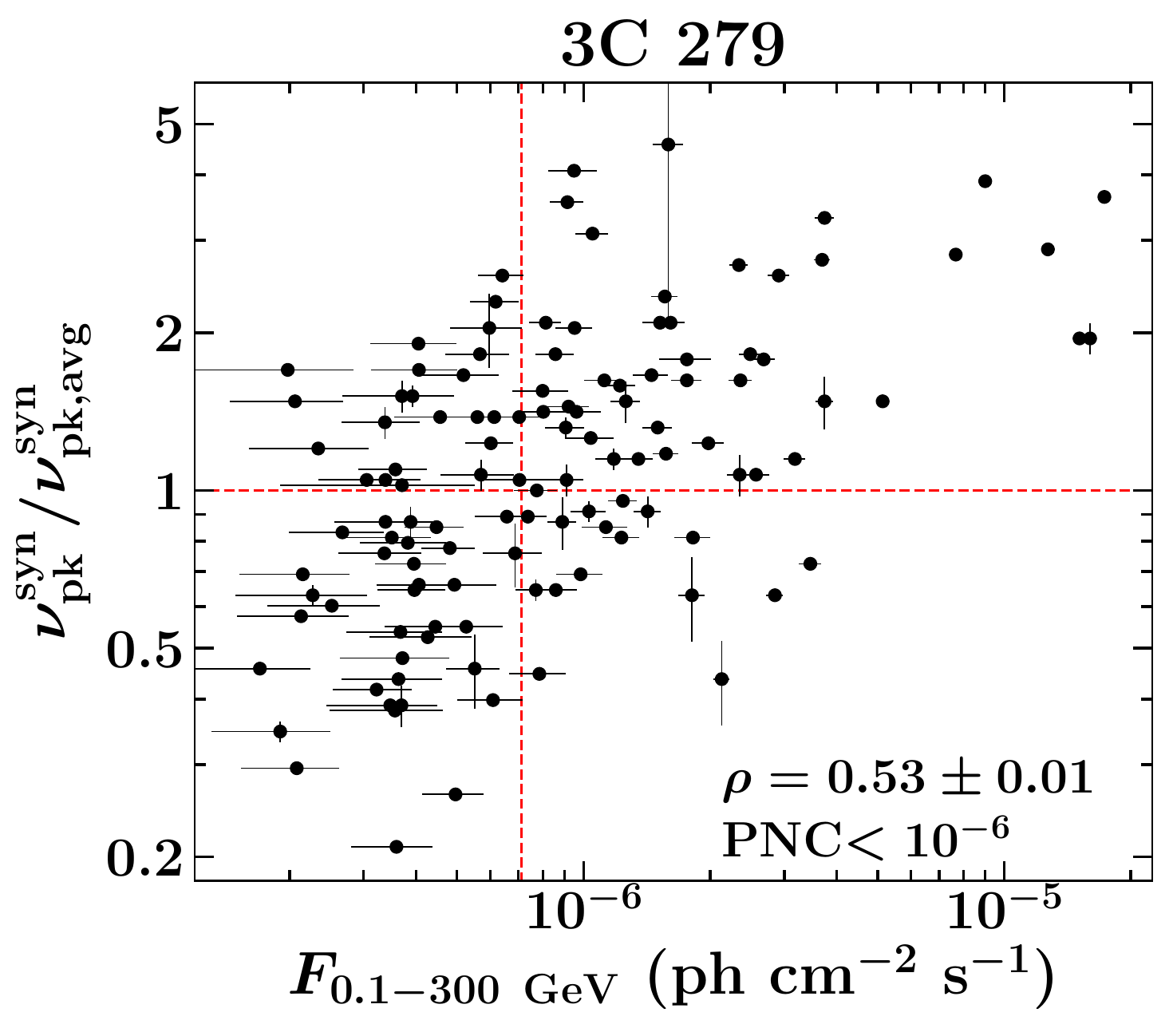} 
\includegraphics[scale=0.36]{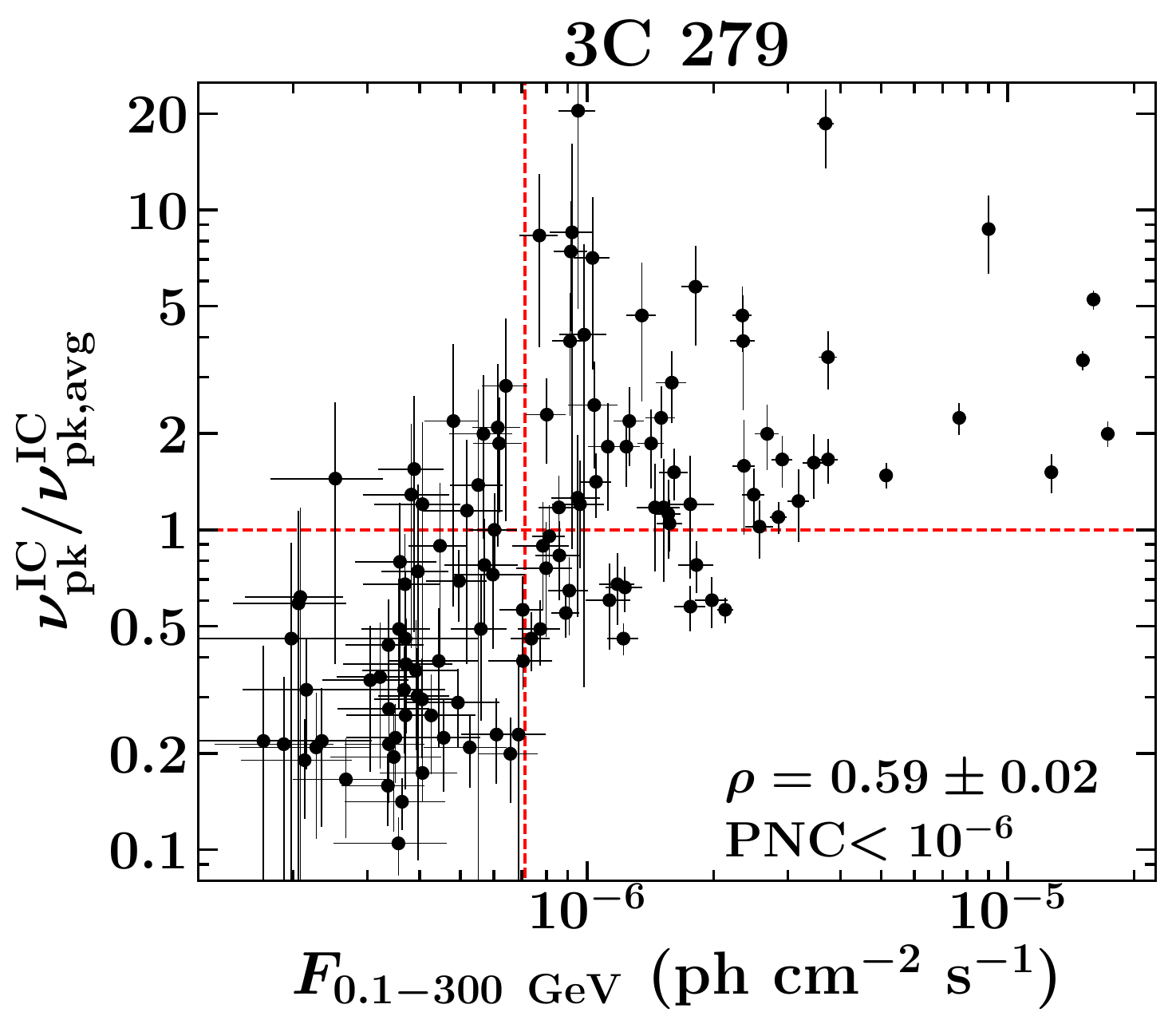} 
\includegraphics[scale=0.36]{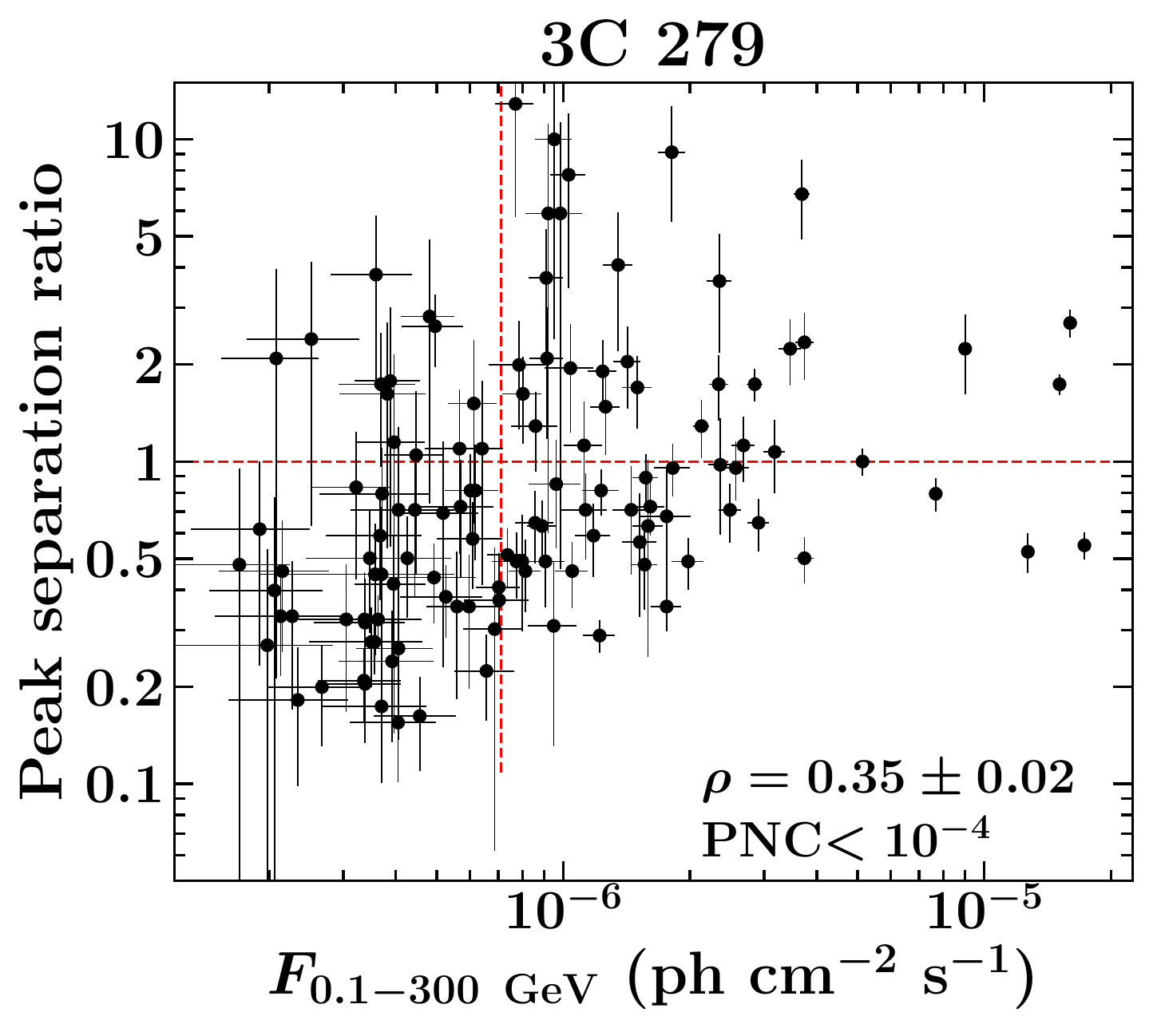} 
}
\hbox{
\includegraphics[scale=0.36]{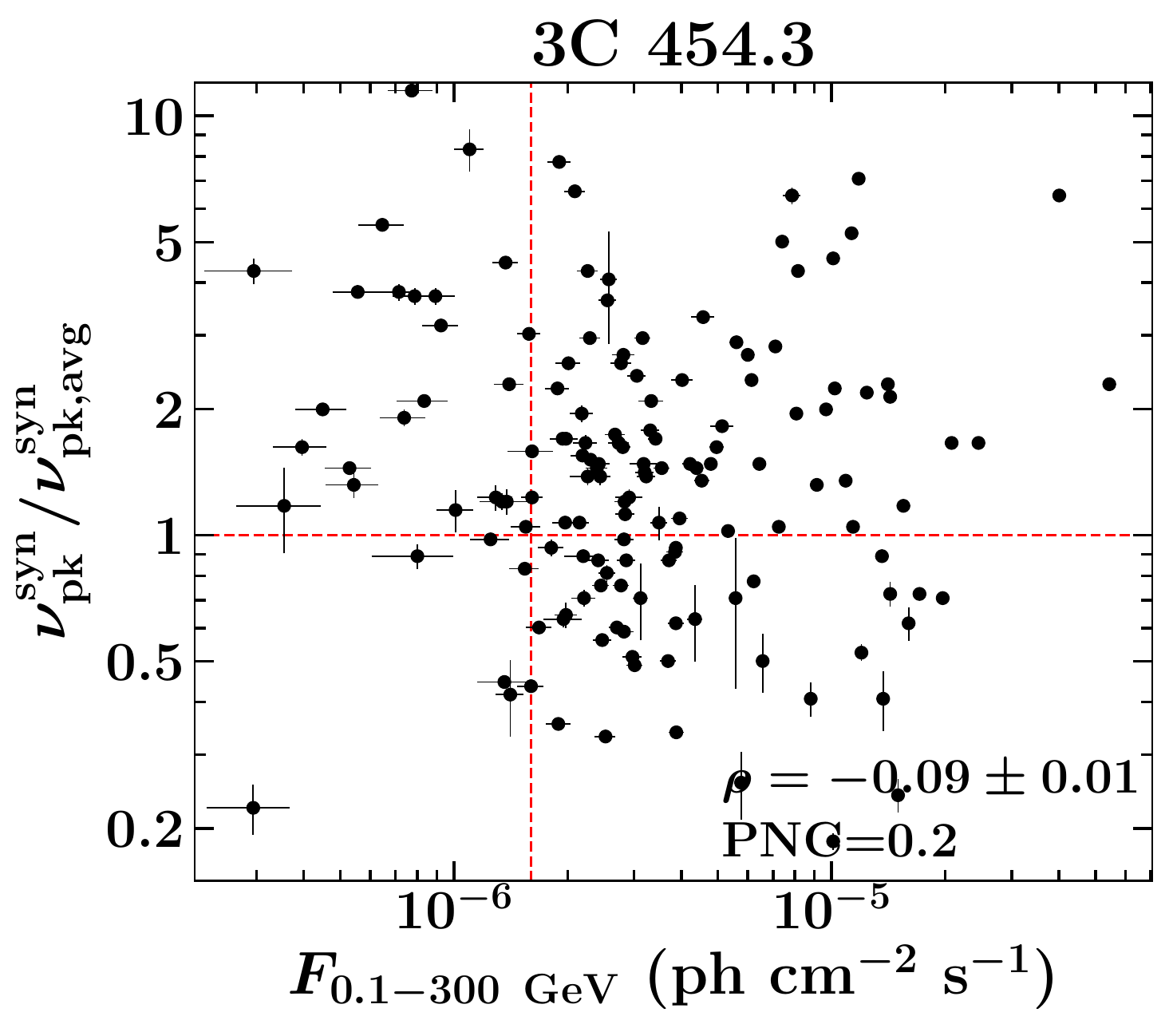} 
\includegraphics[scale=0.36]{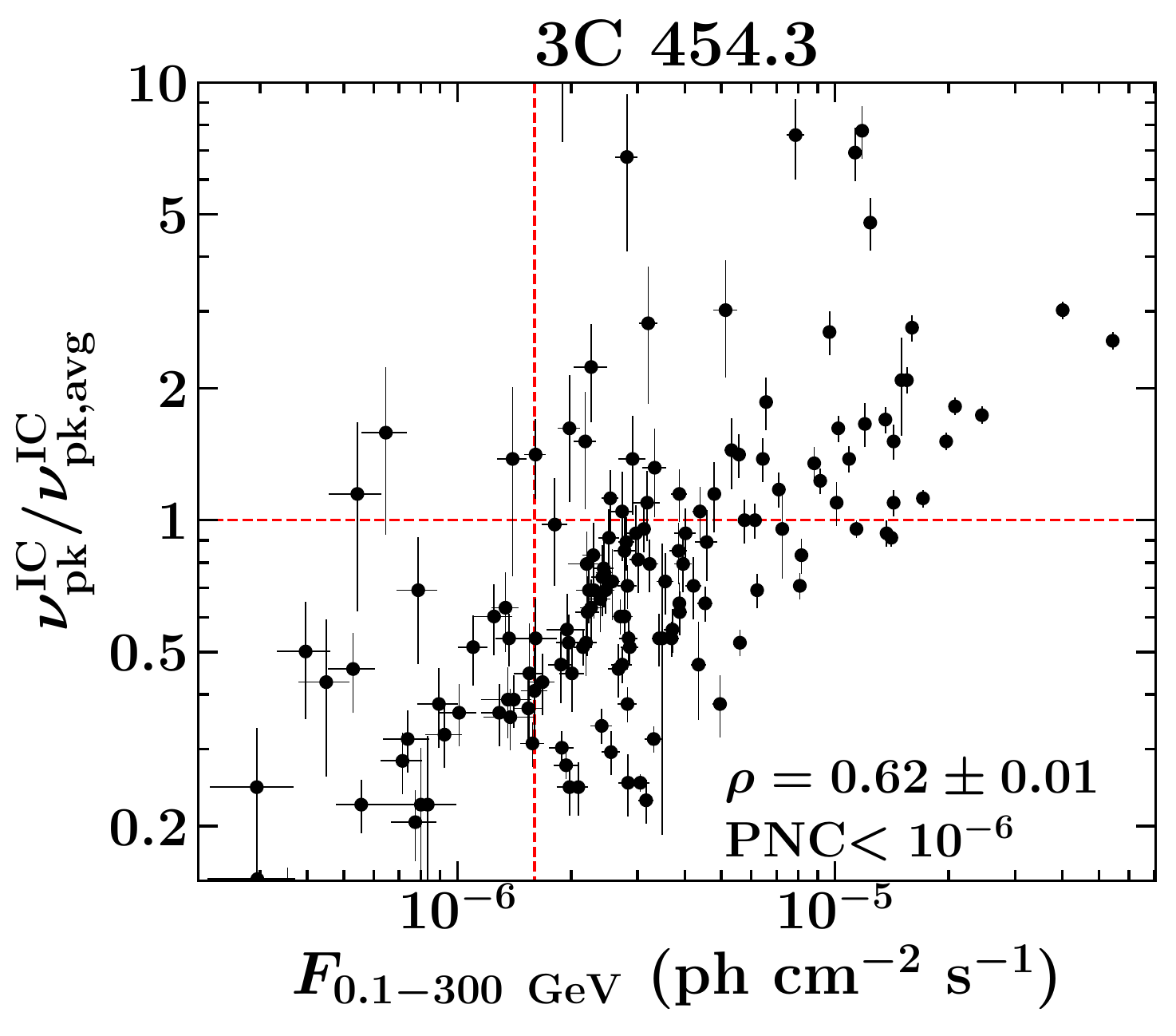} 
\includegraphics[scale=0.36]{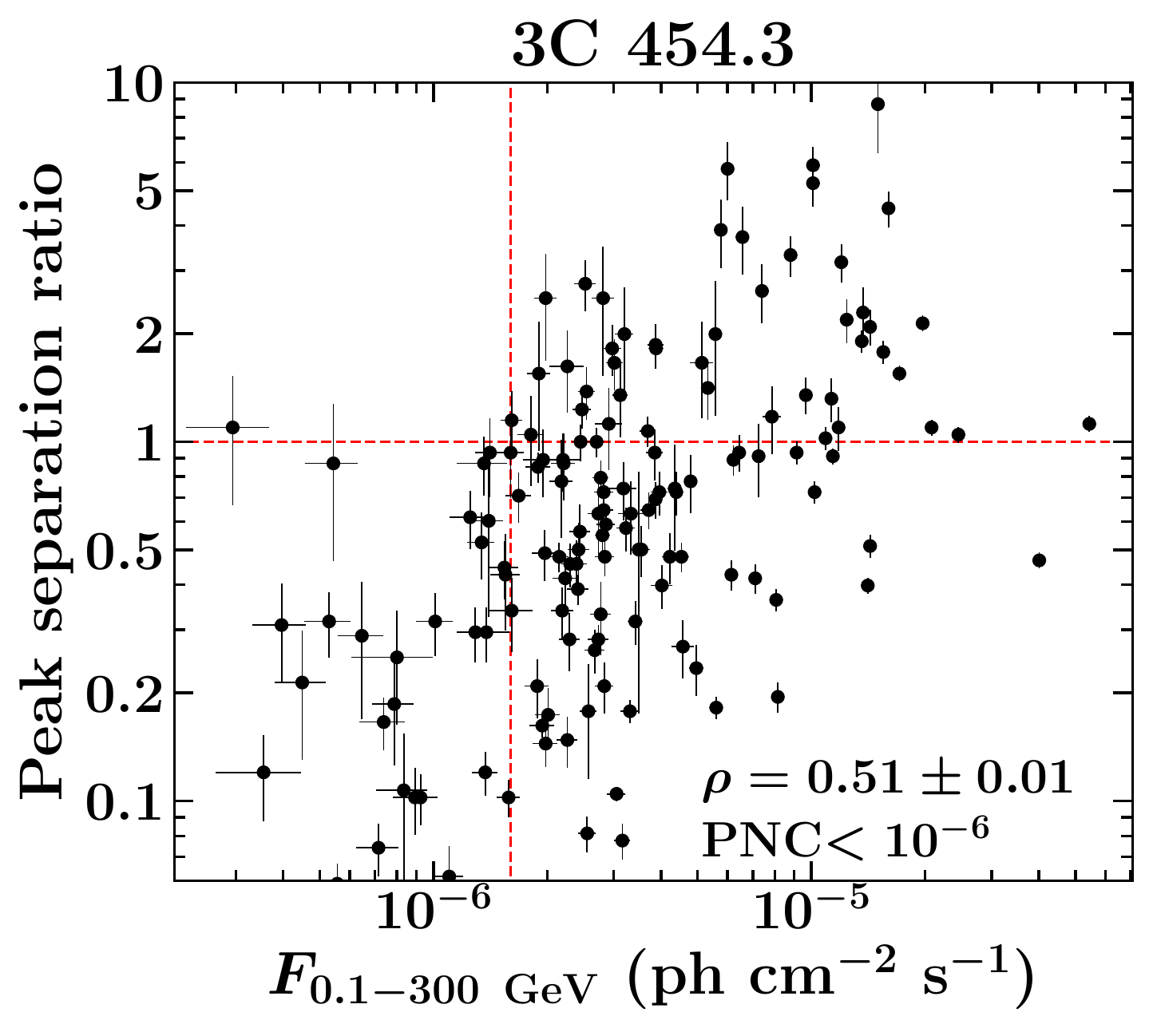} 
}
}
\caption{Variations of $\nu_{\rm pk}^{\rm syn}$ (left), $\nu_{\rm pk}^{\rm IC}$ (middle), and their separation (right) as a function of the \gm-ray flux 
of 3C 279 (top) and 3C 454.3 (bottom). The red dashed lines refer to the average activity state parameter. $\rho$ is the Spearman's rank correlation 
coefficient, and PNC is the probability of the quantities not being correlated. }
\label{fig:peak}
 \end{figure*}

Figure~\ref{fig:peak} shows the variations of the synchrotron peak frequency ($\nu_{\rm pk}^{\rm syn}$), inverse Compton peak frequency 
($\nu_{\rm pk}^{\rm IC}$), and their separation measured during various \gm-ray flux states, normalized to those derived from their average 
activity state SEDs by \citet[][]{2021ApJS..253...46P}. The strength of the correlation of the plotted quantities was determined by computing 
Spearman's rank correlation coefficient ($\rho$) taking uncertainties into account. Both $\nu_{\rm pk}^{\rm syn}$ and $\nu_{\rm pk}^{\rm IC}$ 
have been found to correlate positively with the \gm-ray activity for 3C 279. For 3C 454.3, $\nu_{\rm pk}^{\rm IC}$ follows the same trend 
though $\nu_{\rm pk}^{\rm syn}$ does not show any significant correlation with the \gm-ray flux. Interestingly, the separation of both SED 
peaks strongly correlates with the \gm-ray flux for both sources. This can be understood by comparing the $\nu_{\rm pk}^{\rm syn}$ and 
$\nu_{\rm pk}^{\rm IC}$ panels in Figure~\ref{fig:peak} which reveals that the extent of variation is larger for the latter. In other words,  
$\nu_{\rm pk}^{\rm IC}$ shifts considerably more than $\nu_{\rm pk}^{\rm syn}$ during \gm-ray flares, thus leading to an overall increase 
in the SED peak separation.

\begin{figure*}[t!]
\hbox{\hspace{2.5cm}
\includegraphics[scale=0.4]{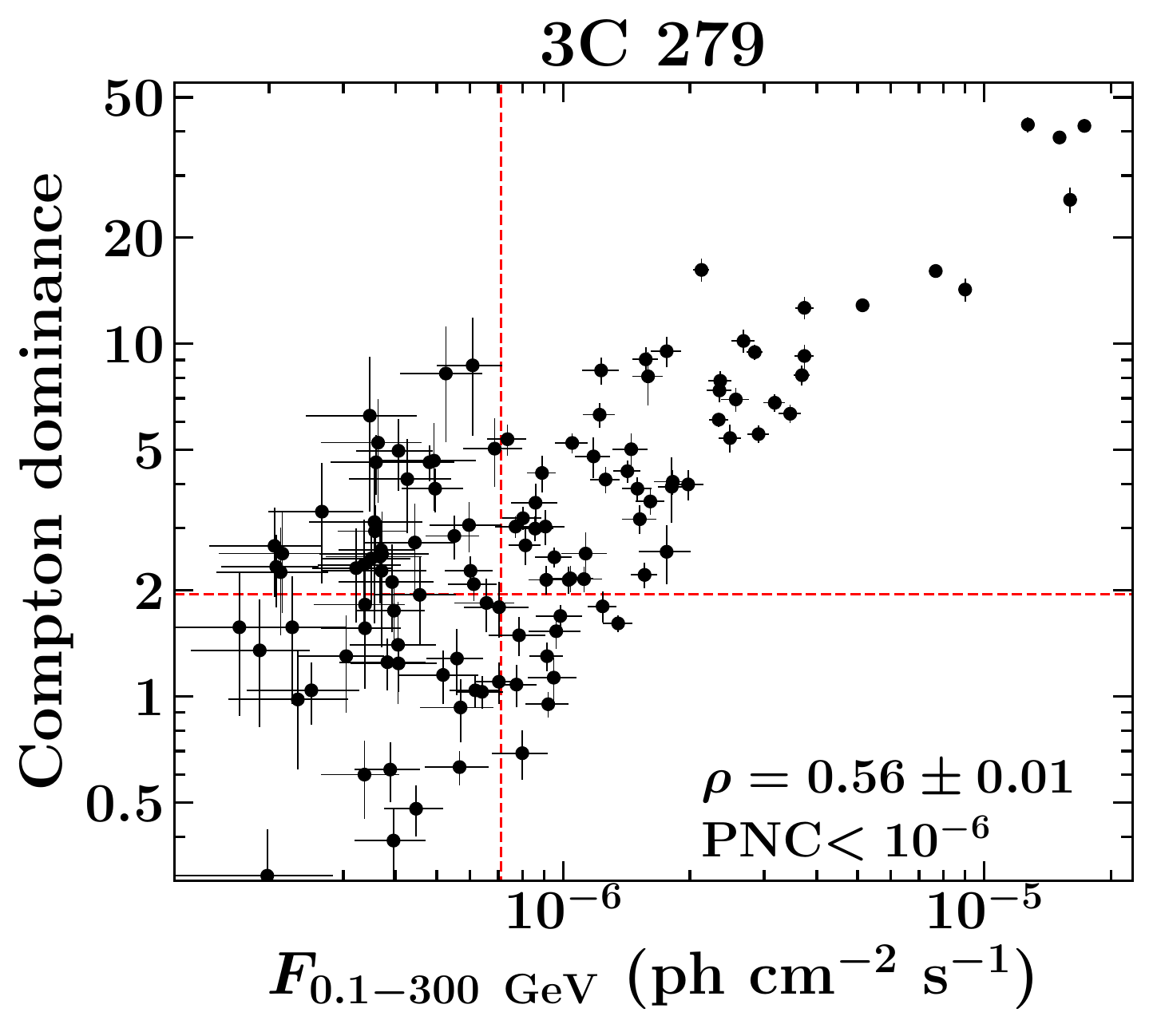} \hspace{0.5cm}
\includegraphics[scale=0.4]{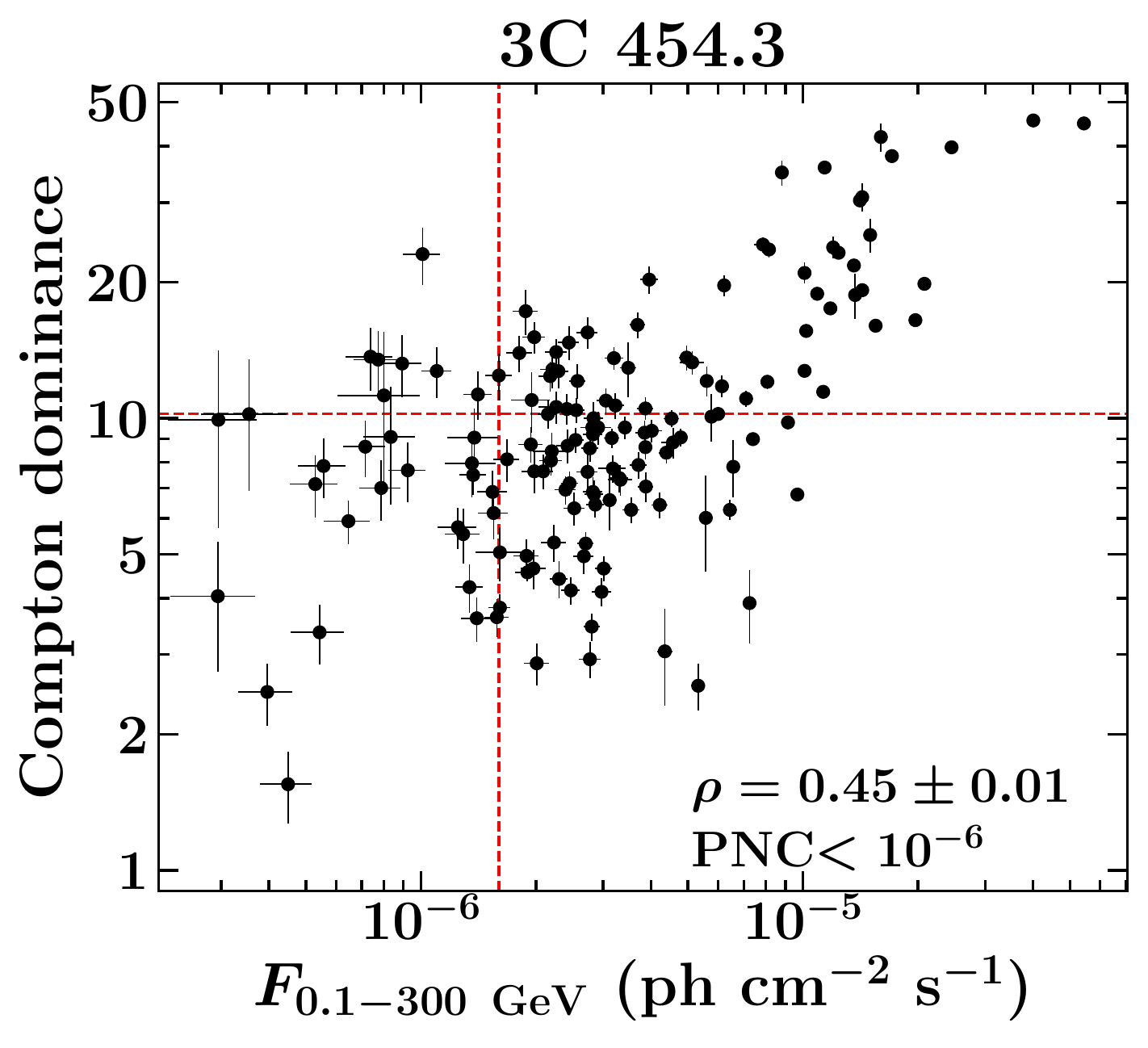} 
}
\caption{Compton dominance as a function of the \gm-ray flux. Other information are same as in Figure~\ref{fig:peak}.}
\label{fig:cd}
 \end{figure*}

We have also derived the ratio of the inverse Compton to synchrotron peak luminosities, also known as Compton dominance, and show it 
as a function of the \gm-ray flux in Figure~\ref{fig:cd}. In both cases, a strong correlation is found.

One possible caveat of this work could be the fact that we have considered near-simultaneous radio data to constrain the synchrotron peak. One can 
argue that the observed radio emission may originate from more extended regions, further down the jet, than the compact one radiating in 
NIR-to-\gm-ray bands due to the synchrotron self-absorption process. Though our analysis method does not rely on any such assumptions, 
we performed the following test to reveal its impact on the results. Since the radio emission produced by the plasma blob radiating 
NIR-to-\gm-rays is expected to be $\leq$ the observed flux, we recomputed $\nu_{\rm pk}^{\rm syn}$ and determine the strength 
of correlations described above by dividing the observed radio flux by 10. No assumption was made on the spectral shape of the 
self-absorbed radio flux. In all cases, the results followed the same trends and the derived correlations remain significant. 
The changes were noticed only in the derived ratios plotted in Figure~\ref{fig:peak} and \ref{fig:cd}. 
For example, decreasing the radio flux reduced the corresponding synchrotron peak luminosity thereby increasing the Compton 
dominance which still showed a strong correlation with the \gm-ray flux, $\rho=0.58\pm0.01$ and $0.43\pm0.01$ for 3C 279 
and 3C 454.3, respectively.

\section{Physical Interpretation}\label{sec:discussion}
The multiwavelength variability and correlation studies of both 3C 279 and 3C 454.3 have been extensively studied, especially during major flares
\citep[e.g.,][]{2009A&A...494..509G,2009ApJ...690.1018V,2011A&A...534A..87R,2012ApJ...754..114H,2013ApJ...773..147J,2015ApJ...803...15P,2016ApJ...826...54D,2016ApJ...832...17B,2020MNRAS.492.3829L,2021PASJ...73..850Z}.

In most works discussing the leptonic modeling of blazar SEDs, the $\gamma$-ray emission from FSRQs is dominated by
inverse-Compton scattering of a target radiation field external to the jet, such as line-dominated optical -- UV radiation
from the Broad Line Region or infrared emission from a dusty torus from a relativistic non-thermal electron population
assumed to be isotropic in the co-moving frame of an emission region moving with bulk Lorentz factor $\Gamma$ along
the jet \citep[e.g.,][]{Ghisellini2010,Boettcher2013}.  This results in relativistic Doppler boosting of the radiation with 
respect to the co-moving frame of the emission region by a factor of 
$\delta = \left( \Gamma \, [1 - \beta_{\Gamma} \cos\theta_{\rm obs}] \right)^{-1}$ in frequency and a factor of 
$\delta^4$ in $\nu F_{\nu}$ peak flux. In such models, therefore, we find that the synchrotron peak is determined
by the source parameters as

\begin{equation}
\nu_{\rm pk}^{\rm syn} \propto B \, \gamma_p^2 \, \delta
\label{nusypeak}
\end{equation}

and

\begin{equation}
\nu F_{\nu, \rm pk}^{\rm syn} \propto N_e (\gamma_p) \, \gamma_p^3 \, B^2 \, \delta^4
\label{nuFnusypeak}
\end{equation}

where $B$ is the magnetic field and $\gamma_p$ is the peak of the electron spectrum in a $\gamma^3
N_e(\gamma)$ representation. In the case of FSRQs with an inverse-Compton peak in the MeV -- GeV
regime, inverse-Compton scattering to photon energies around the peak is likely dominated by Compton
scattering in the Thomson regime. Therefore, the inverse-Compton peak location is determined by

\begin{equation}
\nu_{\rm pk}^{\rm IC} \propto \nu_{\rm ext} \, \gamma^2 \, \Gamma \, \delta 
\label{nuCpeak}
\end{equation}

and

\begin{equation}
\nu F_{\nu, \rm pk}^{\rm IC} \propto u_{\rm ext} \, N_e (\gamma_p) \, \gamma_p^3 \, \Gamma^2 \, \delta^4
\label{nuFnuCpeak}
\end{equation}

where $u_{\rm ext}$ and $\nu_{\rm ext}$ are the energy density and peak frequency of the external radiation field
in the stationary frame of the AGN. 

The peak separation ratio therefore depends on source parameters as

\begin{equation}
{\nu_{\rm pk}^{\rm IC} \over \nu_{\rm pk}^{\rm syn}} \propto {\nu_{\rm ext} \, \Gamma \over B}
\label{nuratio}
\end{equation}

and the Compton dominance as

\begin{equation}
{\nu F_{\nu, \rm pk}^{\rm IC} \over \nu F_{\nu, \rm pk}^{\rm syn}} \propto {u_{\rm ext} \, \Gamma^2 \over B^2}.
\label{CD}
\end{equation}

As evident from Eq. (\ref{nuFnuCpeak}), $\gamma$-ray flaring may be caused by several factors, which we will discuss
in the following: (1.) an increase in the number of relativistic electrons $N_e(\gamma)$; (2.) a hardening of the radiating
electron distribution, resulting in an increase in $\gamma_p$; (3.) an increase in the energy density of the external radiation
field, $u_{\rm ext}$, and/or (4.) an increase in the bulk Lorentz factor $\Gamma$ and Doppler factor $\delta$. 

\begin{enumerate}

\item{A pure increase of the electron number density (without significant spectral changes or other parameter changes) 
will lead to an equal increase of both radiation components (synchrotron and inverse-Compton). It will not shift the peak frequency of 
either component and will not alter the Compton dominance. Hence, such a scenario is ruled out as the dominant cause 
of $\gamma$-ray variability by the significant correlation of both the peak separation and the Compton dominance with 
flux in both sources. }

\item{A spectral hardening of the electron distribution, as expected in most particle acceleration scenarios, such as internal
shocks or magnetic reconnection, shifting $\gamma_p$ to higher values, without any other parameter changes, will lead
an equal increase of the synchrotron and Compton peak frequencies and peak fluxes. In order to increase the peak separation,
other parameter changes would need to be invoked. It has been found in several modeling studies of blazar variability 
\citep[see, e.g.,][]{Zhang2015,BB2019} that a reduction in the magnetic field accompanied by more efficient particle
acceleration can successfully be invoked to reproduce the observed increase of the Compton dominance and peak 
separation. In particular, it may explain the lack of a correlation of the synchrotron peak frequency with the $\gamma$-ray
flux, as we found for 3C454.3. Such a reduction of the magnetic field accompanying $\gamma$-ray flaring activity could
indicate magnetic energy conversion, as in magnetic reconnection, transferring magnetic-field energy into relativistic
particle energy. 

An interesting alternative explanation has recently been suggested by \cite{Sobacchi2019} and \cite{Sobacchi2021}:
Those authors propose that stochastic particle acceleration mediated by magneto-hydrodynamic (MHD) turbulence
will cause the highest-energy particles to maintain small pitch angles with respect to the magnetic field. This will
naturally reduce both the synchrotron peak frequency and the synchrotron flux. It may result in flaring due to enhanced
particle acceleration, shifting $\gamma_p$ to higher values, not leading to a significant increase of the synchrotron
peak, while the inverse-Compton peak frequency and flux may increase significantly. }

\item{Blazar variability models invoking a temporary increase of the energy density of the external target photon field
for inverse-Compton scattering have been proposed in the form of a synchrotron mirror model, in which part of the jet synchrotron
emission is reflected by a cloud near the jet trajectory \citep{BD1999,Tavani2015} or by a stationary jet feature, as
in the ``ring of fire model'' \citep{MacDonald2017}. If the particle acceleration characteristics in the emission region
are not changing in such a scenario, as the emission region passes the mirror, the enhanced radiative cooling is
naturally expected to decrease $\gamma_p$, thus, in fact, leading to a decrease of both the synchrotron and
inverse-Compton peak frequencies as well as a reduction of the synchrotron peak flux. If these were the only changes in such
a scenario, one would expect to see the observed increase in Compton dominance correlated with increasing flux,
but not the increasing peak separation and the positive correlation of the peak frequencies with flux. In order to
reproduce those latter correlations, a reduction of the magnetic field, along with increased particle acceleration
efficiency (increasing $\gamma_p$) would have to occur along with the passing of the mirror. While additional
particle injection and acceleration may be the result of a shock caused by the mechanical interaction of the 
emission region with the obstacle \citep[see, e.g.,][]{Barkov2012,Zacharias2017}, one would expect a compression 
and, thus, increase of the magnetic field to result as well, leading to a decrease of the peak separation. Thus, it
seems difficult to imagine a plausible scenario of an increasing external radiation field to reproduce the flux
correlations found in our study.  }

\item{A change in the bulk Lorentz factor and Doppler factor will lead to enhanced Doppler boosting of both 
the peak frequencies and fluxes of both radiation components. If such a change were effected by a change in
the viewing angle, not changing $\Gamma$, both radiation components would be affected in the same way,
in contrast to the observed peak separation vs. flux and Compton dominance vs. flux correlations. Thus, more
plausible would be an increase of the Doppler factor, which, for initial viewing angles $\theta_{\rm obs} < 1/\Gamma$,
will also lead to an increase of $\delta$. Positive correlations of all quantities (both peak frequencies, peak separation,
and Compton dominance) are expected in this case, as observed for 3C279 \citep[see also][]{2020MNRAS.492.3829L} In order to explain the lack of a correlation
between $\nu_{\rm pk}^{\rm sy}$ and $\gamma$-ray flux for 3C 454.3, a reduction of the magnetic field, as in scenarios 1.
and 2., would need to be invoked.}

\end{enumerate}

\section{Summary}
In this work, we have studied the evolution of the SED peaks as a function of the 
\gm-ray activity of two bright blazars with good multi-wavelength coverage. The separation of the SED peaks and the Compton
dominance are found to strongly correlate with the \gm-ray flux activity for both sources, indicating a similar mechanism to
be responsible for the observed \gm-ray flares. We have interpreted these findings based on the spectral hardening of the radiating
electron population and/or enhancement of the Doppler factor. In both cases, magnetic reconnection may play a major role
in powering the \gm-ray flares. Similar results have been reported in various studies focused on the major flaring events detected from 3C 279 and 3C 454.3 \citep[cf.][]{2011MNRAS.410..368B,2019MNRAS.484.3168S}. Furthermore, these objects have occasionally shown `unusual' SEDs during flaring episodes that might not have been included in our analysis due to lack of near-simultaneous data e.g., at radio or X-ray wavelengths \citep[e.g.,  2013 December flare of 3C 279;][]{2016ApJ...817...61P}. Such peculiar flaring events are crucial in understanding the origin of \gm-ray flares and associated diverse radiative mechanisms by applying various physically-motivated models rather than the polynomial function used in this study. Therefore, the results presented here should be considered as representing the overall broadband spectral behavior of bright \fermi~blazars as a function of their \gm-ray activities. The proposed hypotheses can be tested for a larger sample of blazars with the accumulation of 
contemporaneous radio-to-\gm-ray data made available by the current- and next-generation of telescopes.

\facilities{\fermi-LAT, \swift, SMA, ALMA}

\software{{\tt emcee} \citep{emcee}, XSPEC \citep[v 12.10.1;][]{1996ASPC..101...17A}, Swift-XRT data product generator \citep[][]{2009MNRAS.397.1177E}, fermiPy \citep[][]{2017arXiv170709551W}}

\acknowledgments
We are grateful to the journal referee for constructive feedback.
The work of M. B\"ottcher is supported by the South African 
Research Chairs Initiative (grant no. 64789) of the Department of Science and 
Innovation and the National Research Foundation\footnote{Any opinion, finding 
and conclusion or recommendation expressed in this material is that of the authors 
and the NRF does not accept any liability in this regard.} of South Africa. V.S.P. is grateful to H. Aller and C. Raiteri for providing radio data taken with UMRAO and published 
in \citet[][]{2020MNRAS.492.3829L}, respectively.  This work is partly based on data taken and assembled by the WEBT collaboration and stored in the WEBT archive at the Osservatorio Astrofisico di Torino - INAF (\url{http://www.oato.inaf.it/blazars/webt/}). This paper has made use of up-to-date SMARTS optical/near-infrared light curves that are available at \url{www.astro.yale.edu/smarts/glast/home.php}. Data from the Steward Observatory spectropolarimetric monitoring project were used. This program is supported by Fermi Guest Investigator grants NNX08AW56G, NNX09AU10G, NNX12AO93G, and NNX15AU81G. The Submillimeter Array is a joint project between the Smithsonian Astrophysical Observatory and the Academia Sinica Institute of Astronomy and Astrophysics and is funded by the Smithsonian Institution and the Academia Sinica. This research has made use of data obtained through the High Energy Astrophysics Science Archive Research Center Online Service, provided by the NASA/Goddard Space Flight Center. This research has made use of NASA's Astrophysics Data System Bibliographic Services.

\appendix
The multiwavelength data used in this paper is provided in Table~\ref{tab1} and \ref{tab2} for 3C 279 and 3C 454.3, respectively.

\begin{table}
\caption{The multiwavelength data for 3C 279.\label{tab1}}
\begin{center}
\begin{tabular}{cccc}
\hline
$T_{\rm start}$ & $T_{\rm stop}$ & Frequency & Luminosity \\
~[1] & [2] & [3]  & [4] \\ 
\hline
54685.7	& 54688.7	& 22.402	& 46.949$\pm$0.136\\
54685.7	& 54688.7	& 22.648	& 46.876$\pm$0.103\\
54685.7	& 54688.7	& 22.893	& 46.803	$\pm$0.083\\
54685.7	& 54688.7	& 23.139	& 46.730$\pm$0.088\\
54685.7	& 54688.7	& 23.385	& 46.657$\pm$0.114\\
\hline
\end{tabular}
\end{center}
\tablecomments{The column information are as follows: Col.[1] and [2]: start and stop time (in MJD) of the \fermi-LAT light curve bins in which there is at least one observations taken in the radio, IR$-$UV, X- and \gm-ray bands; Col.[3]: observed log-scale frequency (in Hz); and Col.[4]: observed log-scale luminosity (in \lum). \\
(This table is available in its entirety in a machine-readable form in the online journal. A portion is shown here for guidance regarding its form and content.)}
\end{table}

\begin{table}
\caption{The multiwavelength data for 3C 454.3.\label{tab2}}
\begin{center}
\begin{tabular}{cccc}
\hline
$T_{\rm start}$ & $T_{\rm stop}$ & Frequency & Luminosity \\
~[1] & [2] & [3]  & [4] \\ 
\hline
54682.7	& 54685.6	& 22.402	& 48.502$\pm$0.032\\
54682.7	& 54685.6	& 22.718	& 48.488$\pm$0.017\\
54682.7	& 54685.6	& 23.034	& 48.434$\pm$0.021\\
54682.7	& 54685.6	& 23.350	& 48.341$\pm$0.025\\
54682.7	& 54685.6	& 23.666	& 48.208$\pm$0.035\\
\hline
\end{tabular}
\end{center}
\tablecomments{All information are same as in Table~\ref{tab1}. \\
(This table is available in its entirety in a machine-readable form in the online journal. A portion is shown here for guidance regarding its form and content.)}
\end{table}

\bibliographystyle{aasjournal}

\end{document}